\documentclass[oldversion]{aa}

\usepackage{url}
\usepackage{rotating}
\usepackage{epsfig}
\usepackage{natbib}
\usepackage{amsmath}

\def\apjl{\rm{ApJL}}               % Astrophysical Journal, Letters
\def\apjs{\rm{ApJS}}               % Astrophysical Journal, Supplement
\def\ao{\rm{Appl.~Opt.}}           % Applied Optics
\def\aap{\rm{A\&A}}                % Astronomy and Astrophysics
\begin{document}

\title{Optimal placement of a limited number of observations for period
searches}

\author{Eric S. Saunders \and Tim Naylor \and Alasdair Allan}

\date{Received December 23, 2005; accepted April 13, 2006}
\institute{School of Physics, University of Exeter, Stocker Road, EX4 4QL\\
           \email{[saunders,timn,aa]@astro.ex.ac.uk}}

\abstract{ 
   
   Robotic telescopes present the opportunity for the sparse temporal placement
   of observations when period searching. We address the best way to place a
   limited number of observations to cover the dynamic range of frequencies
   required by an observer. We show that an observation distribution
   geometrically spaced in time can minimise aliasing effects arising from
   sparse sampling, substantially improving signal detection quality. The base
   of the geometric series is however a critical factor in the overall success
   of this strategy. Further, we show that for such an optimal distribution
   observations may be reordered, as long as the distribution of spacings is
   preserved, with almost no loss of quality. This implies that optimal
   observing strategies can retain significant flexibility in the face of
   scheduling constraints, by providing scope for on-the-fly adaptation.
   Finally, we present optimal geometric samplings for a wide range of common
   observing scenarios, with an emphasis on practical application by the
   observer at the telescope. Such a sampling represents the best practical
   empirical solution to the undersampling problem that we are aware of. The
   technique has applications to robotic telescope and satellite observing
   strategies, where target acquisition overheads mean that a greater total
   target exposure time (and hence signal-to-noise) can often in practice be
   achieved by limiting the number of observations.

}


\maketitle

\section{Introduction} \label{Section:Introduction}
   
   %% Advantages of single robotic telescopes.

   Recent years have seen the rise of research-grade robotic telescopes such as
   the Liverpool Telescope \citep{Bode99LT}, as well as a sharp increase in the
   total number of autonomous telescopes active across the world
   \citep{Hessman04GridObserving}. The emergence of these telescopes and
   the flexibility of their operation jusitifies renewed interest in the optimum
   scheduling of observations. Our aim is to discuss in detail one such strategy
   and to demonstrate its optimal properties for our project. Individual
   robotic facilities offer new scheduling possibilities to the observer by
   relaxing the requirement that a human be physically present at the telescope
   at all times. Robotic telescopes offer the potential for ideal sampling of
   periodic signals, since observations may be requested more sparsely and over
   a longer timescale than a traditional continuous observing run.

   %% Advantages of multiple robotic telescopes.
   
   Furthermore, when unmanned telescopes are linked by a communication
   infrastructure, new ways of performing science become feasible
   \citep{Allan04eStarSPIE}. For example, the diurnal sampling problems that
   plague single-site observations of periodic variables can be avoided if an
   observer can acquire data from a second, geographically distant site.
   Continuous monitoring projects such as the Whole Earth
   Telescope \citep{Nather90WET}, and the BiSON network for solar observations
   \citep{Chaplin96BISONPerformance} indicate the success of this approach.

   %% The big question.

   Thus we arrive at a central question in the study of periodic stars: given
   access to round-the-clock observing, how can we best sample a potentially
   periodic star, whose luminosity oscillates with some unknown frequency, such
   that we can minimise the effects of aliasing, while providing a suitably long
   baseline with which to acquire the periods of interest, all while maintaining
   the quality of our data? In particular, how should we choose when to observe
   in those situations where the total number of observations that may be made
   is a limited resource? This situation is commonly the case for robotic
   telescopes, where target acquisition overheads, which include the slew time
   and the CCD readout time, as well as scheduler requirements, can preclude
   large numbers of observations. Although this question reflects our interests
   in using robotic networks to study variable stars, it is equally applicable
   to satellite-based observations.

   The astrophysics of the problem limits the frequencies of interest to an
   astronomer, allowing limits to be placed on the range of frequencies which
   must be correctly reconstructed. Then for an evenly sampled dataset, the
   input signal may be correctly reconstructed if the sampling frequency is at
   least twice the highest frequency $\nu_{\rm max}$ to be detected. Assuming
   the astronomer wishes to see two cycles of any modulation then the lowest
   frequency detectable is given by $2/T$, where $T$ is the duration of  the
   total sampling interval (i.e. the run-length). The value of the required
   sampling frequency $\nu_{\rm N}$ for equally spaced data can in some
   way be viewed as the Nyquist frequency of the dataset, given by 

   \begin{equation}
   \nu_{\rm max} < N / 2T = \nu_{\rm N},
   \label{Equation:Nyquist}
   \end{equation}
   where $N$ is the number of observations. The problem becomes more interesting, however,
   for the case in which there are not sufficient datapoints to fulfill this
   condition. Then the question is whether, under these conditions, we can
   sample in such a way that we can recover a signal of acceptable quality.
   Alternatively, we can reverse the problem: how many datapoints do we need in
   order to recover an acceptable signal? In what follows, we show that there is
   scope for significant optimisation of observation placement in this
   undersampled regime.

\section{Previous work} \label{Section:Previous work}

   %% Motivation: Lack of knowledge.

   In the field of time domain astronomy, there is a relative dearth of
   literature regarding the best way to sample data in time. In contrast, much
   attention has been focussed on the problem of signal extraction from fixed,
   finite datasets \citep[e.g.][]{HorneBaliunas86Sampling,
   Schwarzenberg-Czerny03PeriodSearching}. This is perhaps unsurprising. The
   vast majority of such datasets are acquired in the classical situation of a
   single extended observing run, where an astronomer physically travels to a
   remote observing site, accumulating data continuously each night for the
   duration of the run. In this case the observer has relatively little choice
   about how to space the observations, and normally opts for the safe strategy
   of using all of the available time for continuous observing, acquiring a
   large total number of observations.

   %% Deeming: Summary and clarification. Define window function.

   \citet{Deeming75UnevenSampling} showed that the observed Fourier transform of
   a discrete dataset can be expressed as the convolution of the true Fourier
   transform $F_{\rm N}(\nu)$ with a spectral window $\delta_{\rm N}(\nu)$ that
   fully describes the interference effects between the sampled frequencies,
   such that
   
   \begin{equation}
   F_{\rm N}(\nu) = F(\nu)*\delta_{\rm N}(\nu),
   \end{equation}      
   where $F(\nu)$ is the complex Fourier transform of a function $f(t)$ and is
   defined as

   \begin{equation}
   F(\nu) = \int_{-\infty}^{+\infty} f(t) e^{i 2 \pi \nu t}dt,
   \end{equation}
    and the spectral window is given by
   
   \begin{equation}
   \delta(\nu) = \sum^{N}_{k=1} e^{i2 \pi \nu t_k},
   \end{equation}               
   where $N$ is the total number of observations and $t_k$ is the set of
   observation times. The window function is even, so only positive frequencies
   are evaluated. The physical interpretation of the window function can be seen
   by considering the Fourier transform of an infinite sine function (a delta
   function) convolved with the window function, leading to the observed Fourier
   transform. Adjusting the set of observation times $t_k$ alters the
   window function, and hence the Fourier transform.
   \citet{Deeming75UnevenSampling} showed that aliasing effects in the observed
   Fourier transform arising from the window function could be mitigated by
   introducing irregularities into the sampling pattern. By careful choice of
   sampling it is possible to reconstruct higher frequencies than $\nu_{\rm N}$,
   the highest frequency that could be accurately reconstructed if the same
   number of observations were equally spaced.

   %% Deeming: Limitations.
   
   \citet{Deeming75UnevenSampling} illustrated the behaviour of the window
   function through the use of a simple single-variable model for the $N=25$
   case, empirically deriving a form for the spacing based on visual inspection
   of the resulting spectral window function.
   \citet{MouldEtAl00HSTKeyProjectXXI} adopted a pseudo-geometric spacing as a
   way to maximise the uniformity of their phase spacing, but no details of the
   implemented optimisation strategy were provided. 
   
   The Rayleigh resolution criterion may be written as $\Delta \nu \Delta T
   = 1$. It means that observations of length $\Delta T$ yield the frequency of
   an oscillation with an error of $\Delta \nu=1/ \Delta T$, at worst. After an
   additional time interval $\alpha \Delta T$ this yields an error in the cycle
   count $\Delta \nu \alpha \Delta T = \alpha$. Thus to keep the cycle count
   error less than $\alpha$, subsequent observations ought to be scheduled no
   later than $(1+ \alpha) \Delta T$ from the start. Recurrent and rigid
   application of this known principle yields a geometric series scheduling (we
   are grateful to the referee for suggesting this line of argument).

   We shall show that the choice of geometric base is crucial for such a
   strategy to succeed. The general question of what form an optimal uneven
   spacing should take, when the number of observations and the length of the
   observing run are free parameters remains, to our knowledge, unanswered.

   %% Our result: Preview.
   
   Here we present a detailed analysis of the application of a geometric spacing
   technique to the undersampling problem. Although our sampling is not an
   analytic solution to the general problem of arbitrary placement of the
   dataset $t_k$, it outperforms both the model of
   \citet{Deeming75UnevenSampling} and pure random distributions, and thus
   represents the best current heuristic solution to the sampling problem. In
   Section \ref{Section:Method}, we present the sampling strategy and quality
   metrics by which it is judged. In Section \ref{Section:Results} we compare
   the geometric sampling to the sampling of \citet{Deeming75UnevenSampling} and
   to random sampling and show that noise has no impact on the sampling
   strategy. We also show that the window function retains similar
   characteristics under random \emph{reordering} of  the spacings, an important
   result for practical scheduling scenarios. In Section \ref{Section:Practical
   application} we present a graph indicating optimal geometric samplings for
   observing runs spanning the range from 10 to 500 observations. We present our
   final conclusions in Section \ref{Section:Conclusions}.

\section{Method} \label{Section:Method}

   %% Define our geometric series.

   We generate our observation times with a geometric distribution using the
   scheme

   \begin{equation}
      t_k = \frac{x^k - 1}{x^{(N-1)} - 1} T,
      \label{Equation:Geometric series}
   \end{equation}   
   where $t_k$ is the (time) value of the point in the series, $x$ is the base
   of the series, $k$ is the number of the current data point, $N$ is the total
   number of observations and $T$ is the duration of the total observing
   interval. This produces a geometrically spaced distribution such that 
   $0 \le t_k \le T$.

   The parameter $x$, which we call the \emph{geometric base}, may be
   arbitrarily chosen. Since the window function is in general complex, for
   comparison with \citet{Deeming75UnevenSampling} we follow the convention of
   examining the normalised amplitude $A(\nu)$, defined as the square of the
   amplitude of the window function, normalised by the square of the number of
   datapoints, such that 

   \begin{equation}
   A(\nu) = \frac{|\delta(\nu)|^2}{N^2}.
   \end{equation}   
   We define the \emph{optimal base}, $x_{\rm opt}$, as the value for which the
   overall degree of structure present in $A(\nu)$ is minimised. We use the root
   mean square (RMS) of $A(\nu)$ as a straightforward metric for identifying
   structure. Minimising the RMS thus represents the desire of the astronomer to
   achieve relatively even sensitivity to periods. 

   In order to retain sensitivity to the higher frequency structure we evaluate
   $A(\nu)$ for frequencies in the range $0.1 < \nu_{\rm N} \le 5$ (in units of
   the Nyquist frequency for an equivalent even sampling). In practice, the
   range of the window function that needs to be considered will vary depending
   on the degree of undersampling of a given dataset (which depends on the
   number of observations and the length of the dataset), and the maximum
   frequency to be detected. Since what matters is the frequency with respect to
   the total time interval, we define the dimensionless \emph{relative
   frequency} as 
   
   \begin{equation}
      \nu_{\rm rel} = \nu T .
      \label{Equation:rel_freq}
   \end{equation}   
   Then the relative limiting frequency in units of the relative Nyquist
   frequency is simply $\nu_{\rm max} / \nu_{\rm N}$.

\section{Results} \label{Section:Results}

   \subsection{The best sampling}

   %% FIG: Deeming best (alpha = 1.0) w. fn.
   
   \begin{figure}  
   \vspace{70mm}
      \includegraphics{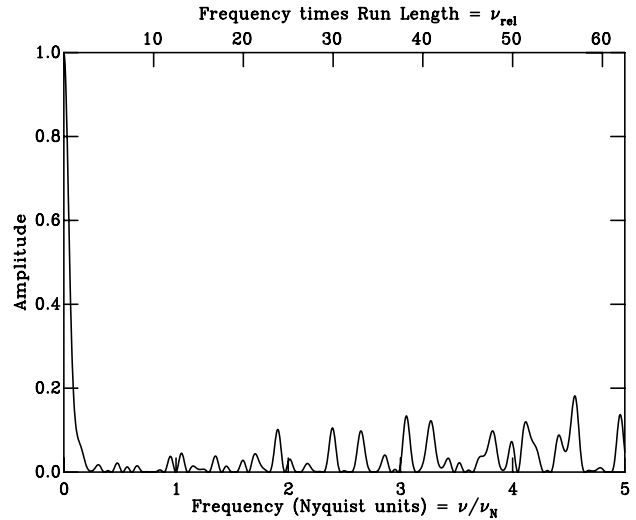}
      \caption{The window function (where the amplitude is given by
      $|\delta(\nu)|^2 / N^2$) produced by Deeming's distribution, with a
      close to optimal value of $\alpha=1.0$. The RMS is 0.0361.
      \label{fig:deeming1.0} }
   \end{figure}

   %% FIG: Random w. fn.

   \begin{figure}  
   \vspace{70mm}
      \includegraphics{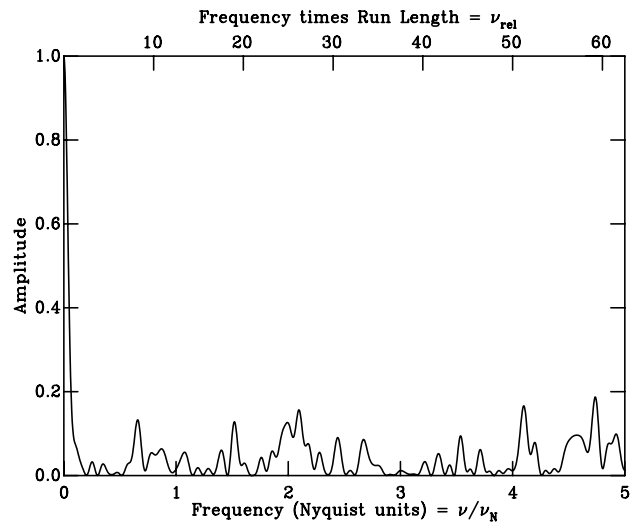}
      \caption{The window function for a typical random distribution of 25 data
      points. The RMS is 0.0386.\label{fig:random_n25} }
   \end{figure}

   %% FIG: Ordered geometric for x=1.124.
   
   \begin{figure}  
   \vspace{70mm}
      \includegraphics{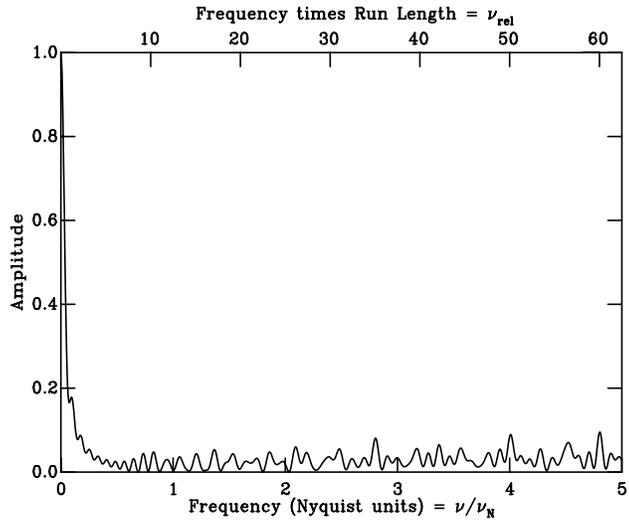}
      \caption{The window function for the optimum geometric sampling of 25 data
      points, for $x=1.124$. The RMS is 0.0201.\label{fig:geometric_n25_x1.124} }
   \end{figure}

   %% FIG: RMS plot for n=25.

   \begin{figure}  
   \vspace{70mm}
      \includegraphics{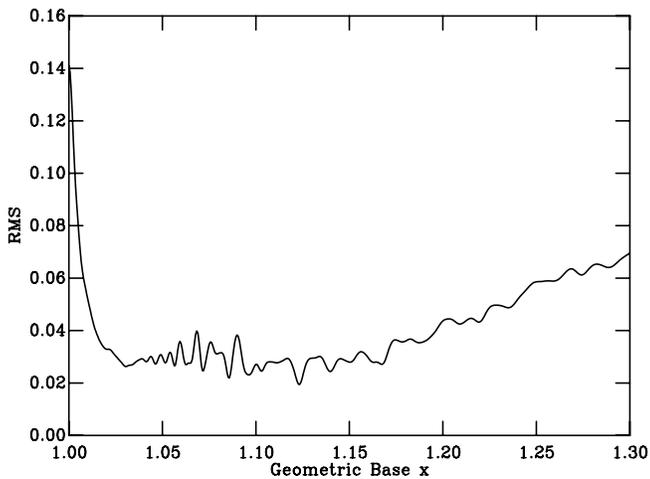}
      \caption{RMS as a function of geometric base, for $N=25$. The optimum
      value is $x=1.124$. \label{fig:geometric_n25_rms} }
   \end{figure}

   %% Discuss figures. Present details of RMS of different plots.

   \citet{Deeming75UnevenSampling} generated 25 datapoints according to the
   formula
   
   \begin{equation}
   t \sim
      \begin{cases}
      k^{-1/ \alpha} & \text{$(k = 1 \ldots 12)$} \\
      (25 - k)^{-1/ \alpha} & \text{$(k = 13 \ldots 24)$}.
      \end{cases}
   \end{equation}                  
   Figure \ref{fig:deeming1.0} is the window function for Deeming's distribution
   with $\alpha=1.0$ - a near-optimal value by visual inspection and RMS
   comparison. Figure \ref{fig:random_n25} presents the window function for a
   randomly spaced set of 25 datapoints. The typical size of the interference
   structures at high frequency are comparable, as indicated by the similar RMS
   values of the two graphs. Figure \ref{fig:geometric_n25_x1.124} is the window
   function for an optimal geometric spacing, where the value of $x$ was chosen
   according to the minimum of Figure \ref{fig:geometric_n25_rms}. The geometric
   spacing exhibits much better high frequency suppression, with an RMS roughly
   half that of either of the other spacings, at the price of a slightly broader
   true peak (centred by construction at $\nu=0$). This broadening is a
   consequence of the undersampling. Although we can push the effective limit
   for frequency reconstruction to higher frequencies, we sacrifice some of our
   knowledge of lower frequencies in order to do so. In particular, the precise
   frequency of the true signal is less clearly represented in the window
   function.

   %% Explain that choice of geometric base is essential.
   
\begin{figure}  
      \vspace{82mm}     
      \includegraphics{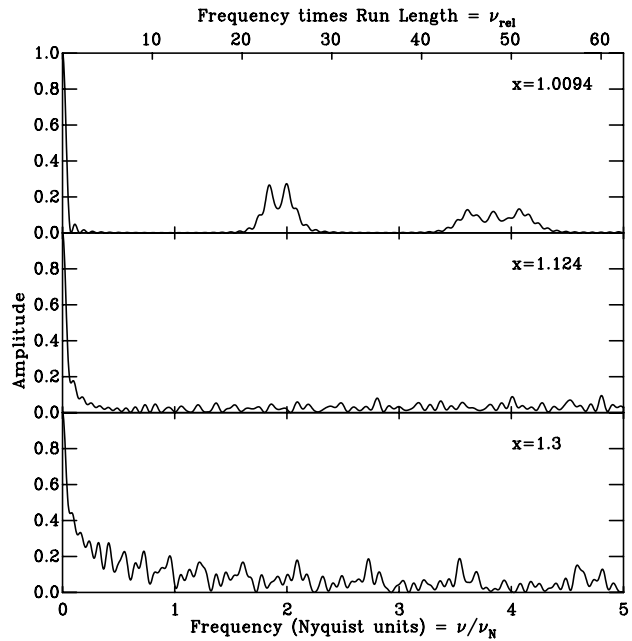}
      \vspace{5mm}
      \caption{Window functions for three geometric spacings of 25 datapoints.
      The top panel illustrates aliasing effects at a near linear sampling of
      $x=1.0094$, while at $x=1.3$ (bottom panel) the effects of near
      frequency interference are apparent. The optimum geometric spacing for
      this number of datapoints is shown for comparison ($x=1.124$, centre panel).
      \label{fig:3geometric} }
   \end{figure}

\begin{figure}  
   \vspace{70mm}
      \includegraphics{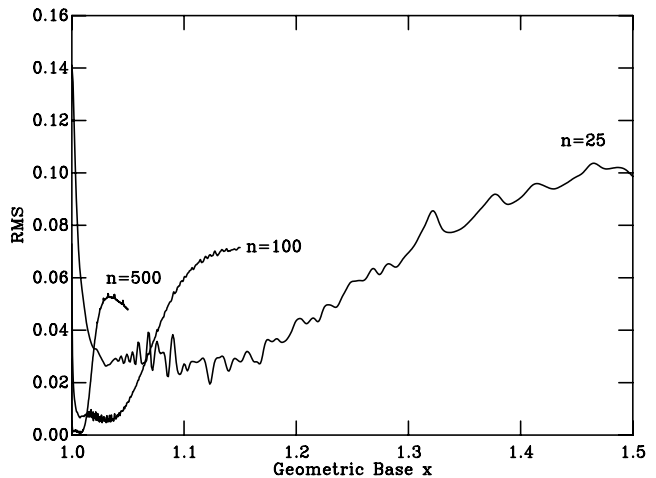}
      \caption{Comparison of the variation of RMS with geometric base, for 
      three dataset sizes. When $n=25$, a relatively broad range of bases lie
      close to the minimum RMS. For larger numbers of datapoints, the
      choice of optimal base is much more tightly constrained. 
      \label{fig:rms_comparison} }
   \end{figure}

   Note that the choice of geometric base is critical to the success of the
   sampling. Figure \ref{fig:3geometric} shows how structure can be reintroduced
   into the window function by a poor choice of geometric base. This situation
   occurs because a sub-optimal geometric base essentially `wastes' information,
   oversampling some frequencies and undersampling others. In the pathological
   case of even sampling, every sampled frequency is a multiple of the true
   frequency, leading to massive redundancy. For datasets with more than 25
   datapoints, the correct choice of geometric base is much more important.
   Figure \ref{fig:rms_comparison} plots RMS as a function of geometric base for
   datasets with 25, 100 and 500 observations. The larger datasets possess much
   narrower minima `valleys' than the $n=25$ dataset. At these dataset sizes,
   any substantial deviation from the optimal base will lead to rapid
   degradation of the window function.

   %% FIG: Wfns for 3 different geometric bases at n=25, highlighting 
   %% differences.

   %% FIG: Comparing three datasets (N=25, 100, 1000) to illustrate how the 
   %% range of optimal bases varies with dataset size.

   \subsection{Noise}

   %% Effect of noise. Note independence of w. fn. from both signal and noise.

   Whilst in the idealised case without noise a suitable geometric sampling is
   the correct sampling, we must consider the effects of noise on the result. If
   noise dominates the periodogram, then there may be little point in optimising
   in the manner described. A range of equally plausible `optimal' geometric
   bases will exist, since the quality of the final observed Fourier transform
   is noise-limited rather than window-function limited.

   When making actual observations, an observer must decide how to divide a
   fixed total exposure time. This choice impacts signal-to-noise, since for
   most practical observing scenarios the signal-to-noise increases as the
   square root of the total exposure time. However, the window function has no
   noise dependency. The quality of a specific window function as a function of
   frequency is determined entirely by the choice of sampling time. This means
   that noise will only be of practical concern to the observing strategy if it
   is greater than the mean amplitude of what \citet{Deeming75UnevenSampling}
   refers to as `near-frequency interference', and which we shall term
   \emph{spectral leakage}.

   % Description of FT grayscale simulations.

   To demonstrate the practical effects of noise on our ability to recover a
   signal from a set of observations, we generate a set of artificial
   lightcurves, each a sinusoid of a single frequency, that span the range from
   1-100 cycles per observing run. We define the \emph{relative period} as
   the ratio of the period $P$ to the observing run $T$, such that 
   
   \begin{equation}
      P_{\rm rel} = \frac{P}{T} \equiv \frac{1}{\nu_{\rm rel}}.
   \end{equation}
   Thus, we consider the range $0.01 \le P_{\rm rel} \le 1$. For each generated
   period, we calculate the Scargle periodogram 
   \citep{Lomb76LeastSquares,Scargle82TimeSeriesII}, as formulated by
   \citet{HorneBaliunas86Sampling}.

   % Definition of the A metric.
   
\begin{figure}  
   \vspace{70mm}
      \includegraphics{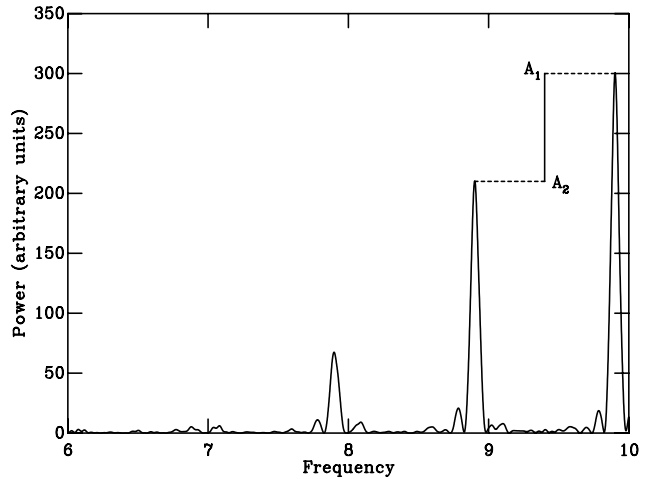}
      
      \caption{Periodogram illustrating how the \emph{A} metric can be applied
      as a measure of relative alias strength. $A_1$ and $A_2$ are the
      amplitudes of the true peak (in this case, the highest peak) and the
      highest \emph{other} peak in the transform. The ratio $\frac{A_1}{A_2}$ is
      our \emph{A} metric. \label{fig:a_alias_example}}

   \end{figure}

\begin{figure}  
   \vspace{70mm}
      \includegraphics{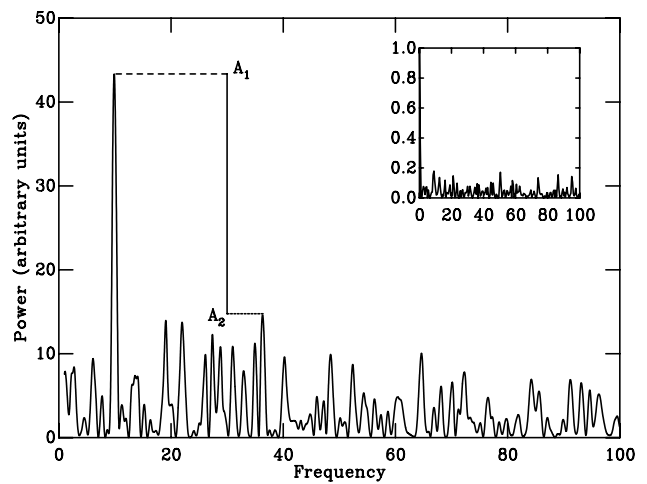}
      
      \caption{Periodogram illustrating how the \emph{A} metric, defined as the
      ratio $\frac{A_1}{A_2}$, can represent the power at the true
      period relative to the spectral leakage. No noise has been
      added to this signal; interference effects are due to the (relatively
      poor) choice of sampling. The signal is a simple sinusoid with
      $P_{\rm rel} = 0.10163$, sampled
      with 100 datapoints using a geometric base of $x = 1.10$, randomly
      respaced. The inset gives the amplitude of the corresponding window
      function, as a function of frequency.
      \label{fig:a_interference_example}}

   \end{figure}

\begin{figure}  
   \vspace{70mm}
      \includegraphics{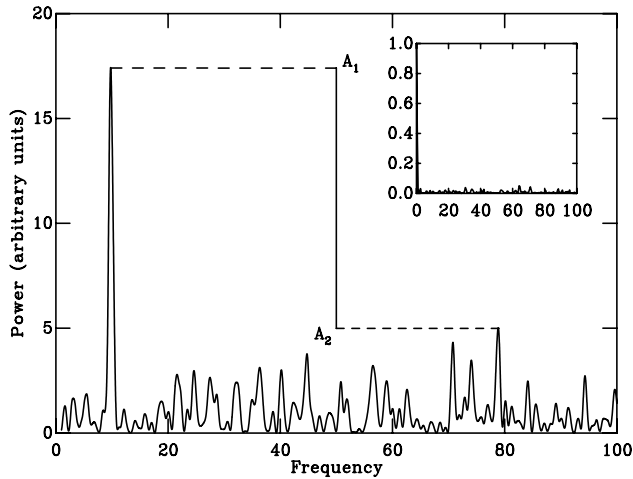}
      
      \caption{Periodogram illustrating how the \emph{A} metric, defined as the
      ratio $\frac{A_1}{A_2}$, can represent the power at the true period
      relative to the overall level of noise present in the signal. The
      input signal is the same as for Fig. \ref{fig:a_interference_example}, but
      with the additional application of noise equal to half the amplitude of
      the signal \label{fig:a_noise_example}. In this case, the geometric
      sampling base used was $x = 1.03$, to minimise the effects of spectral
      leakage. The inset gives the amplitude of the corresponding window
      function, as a function of frequency; note the relatively minor
      interference contributions.}

   \end{figure}

   The quality of the periodogram is assessed by calculating the \emph{peak
   ratio}, $A$, which we define to be the ratio of the powers of the peak
   corresponding to the true period and the highest other peak in the
   periodogram. There are three regimes of interest for this metric. Figure
   \ref{fig:a_alias_example} illustrates the `alias' regime, where the metric
   indicates the extent to which power has been lost to other strong signal
   candidates. When aliases have been largely suppressed or when the overall
   level of noise in the periodogram is large, the metric instead describes the
   prominence of the peak relative to the spectral leakage, or to the
   background noise. These two situations are illustrated in Fig.
   \ref{fig:a_interference_example} and \ref{fig:a_noise_example} respectively.
   The relationship to the alias property or peak amplitude leads us to denote
   this metric \emph{A}.

   %% Period accuracy 'fuzziness'.
   
   The accuracy to which a period detection is required changes \emph{A}. An
   observer would not be interested in all nearby peaks or sub-peaks, but the
   scale on which structure in the periodogram is important is obviously highly
   dependent on the nature of the science programme. For all our simulations, we
   consider periods within 10 per cent of the true period to be adequate, in
   line with typical accuracy requirements for rotation rate histograms
   \citep{Littlefair05IC348}. We assume that our periods have an infinite
   coherence length, i.e. that the phase information does not change over time.

   %% FIG x 3: A example regimes
   
   %% A 'alias' example

   %% A 'interference' example

   %% A 'noise' example

   %% Explain what our noise means.
    
\begin{figure}              
      \epsfig{file=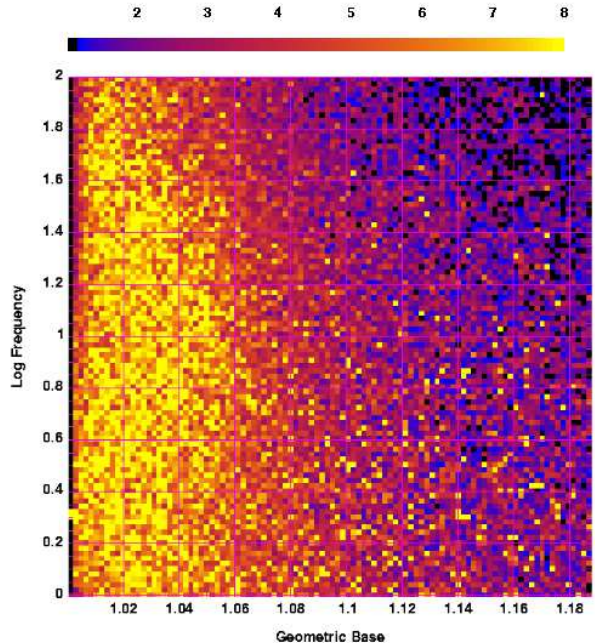, height=\columnwidth}
      \caption{\emph{A} for an $N=100$ ordered geometric lightcurve with noise
       equal to 30\% of the signal amplitude, as a function of geometric base and 
       injected signal frequency.\label{fig:a_n100_noise30} }
   \end{figure}

\begin{figure}       
      \epsfig{file=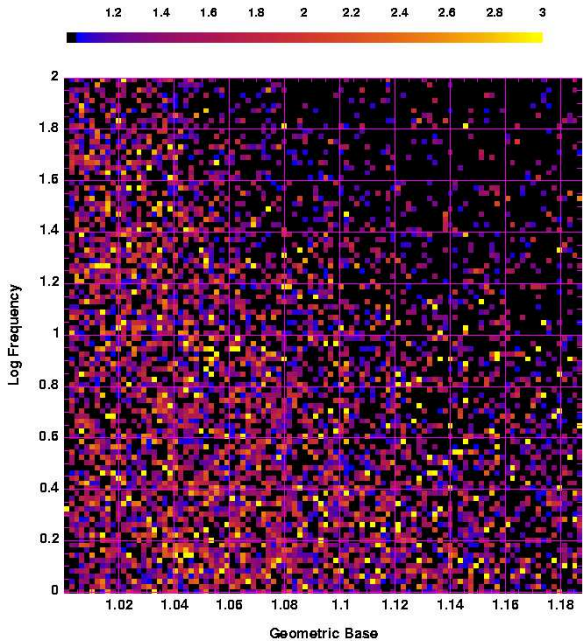, height=\columnwidth}
      \caption{\emph{A} for an $N=100$ ordered geometric lightcurve with noise 
      equal to 100\% of the signal amplitude, as a function of geometric base
      and injected signal frequency. High noise dominates the periodogram,
      washing out structure in the plane. \label{fig:a_n100_noise100} }
   \end{figure}

   For a 100 observation dataset, we apply noise equal to one third of the full
   amplitude of the signal (Figure \ref{fig:a_n100_noise30}) and noise equal to
   the amplitude of the signal itself (Figure \ref{fig:a_n100_noise100}). This
   yields a signal-to-noise of 3.33 and 1 respectively. It should be noted that
   this is not the signal-to-noise normally quoted by astronomers, where the
   time-varying signal is superimposed upon a constant flux. For
   comparison, in a typical observing scenario where the full amplitude of the
   signal modulation might make up around 10 per cent of the incident signal
   flux, our simulated noise would correspond to an observational combined
   signal-to-noise of around 33 and 10.

\begin{figure}  
   \vspace{70mm}
      \includegraphics{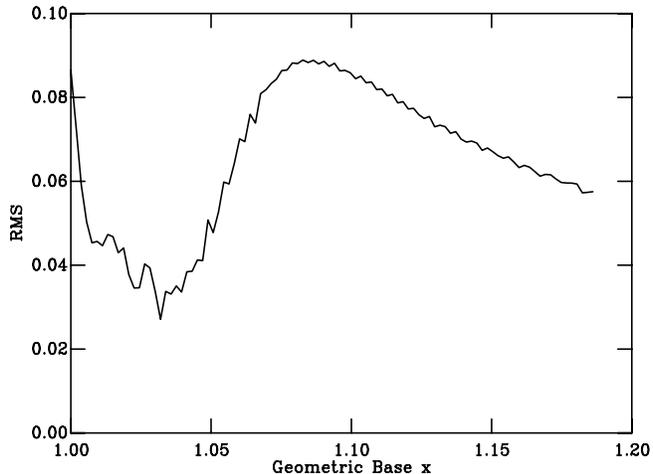}
      \caption{RMS as a function of geometric base, for $N=100$. The optimum
      value is $x=1.032$. \label{fig:geometric_n100_rms} }
   \end{figure}

   In Figure \ref{fig:a_n100_noise30} the noise level is comparable to the
   spectral leakage, and we see the optimal base choice clearly picked out
   by the $A$ metric. For comparison, Figure \ref{fig:geometric_n100_rms} gives
   the RMS of the window function as a function of geometric base for the same
   dataset. Figure \ref{fig:a_n100_noise100} illustrates the degradation of the
   periodogram under the effects of extreme noise. Under these conditions, the
   choice of geometric base is much less important, because the amplitude of the
   noise is typically much greater than the amplitude of the spectral
   leakage.

   %% Optimisation of the sampling strategy remains useful even in the presence
   %% of substantial noise.

   We conclude that noise is not a contributing factor to the choice of optimal
   base for standard astronomical applications.

   %% FIG: A, N=100, noise = 30

    %% FIG: RMS plot for n=100.

   %% FIG: A, N=100, noise = 100

   \subsection{Randomised geometric spacing}

   %% Explain that positions of datapoints may be reshuffled in time, as long
   %% as spacings are preserved.
   
   Given the favourable properties of a standard geometric sampling, we can ask
   whether the order of the sampling matters. We can preserve our choice of
   spacing while modifying the order in which observations take place. This is
   equivalent to shuffling the gaps between observations. One motivation for
   doing this is that it allows individual observations to be placed with much
   more flexibility than an ordered geometric spacing, greatly easing scheduling
   problems.

   %% Describe results.
   
\begin{figure}  
   \vspace{70mm}
      \includegraphics{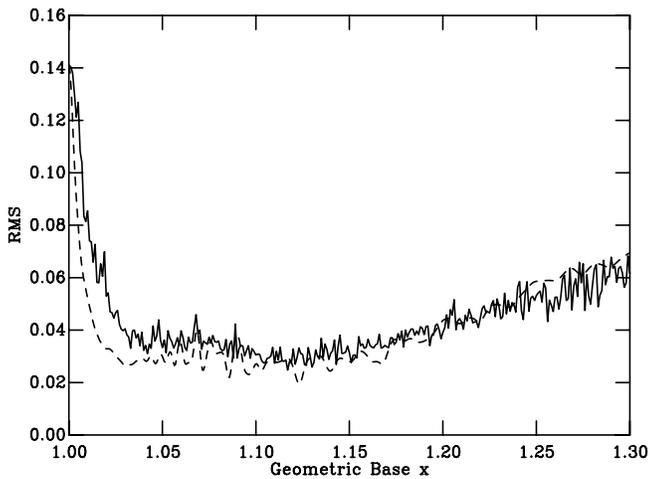}
      \caption{RMS as a function of randomised geometric base, for $N=25$ (solid
      line). For
      comparison, the ordered geometric spacing has been overplotted (dashes).
      \label{fig:rand_geometric_n25_rms} }
   \end{figure}

\begin{figure}  
   \vspace{70mm}
      \includegraphics{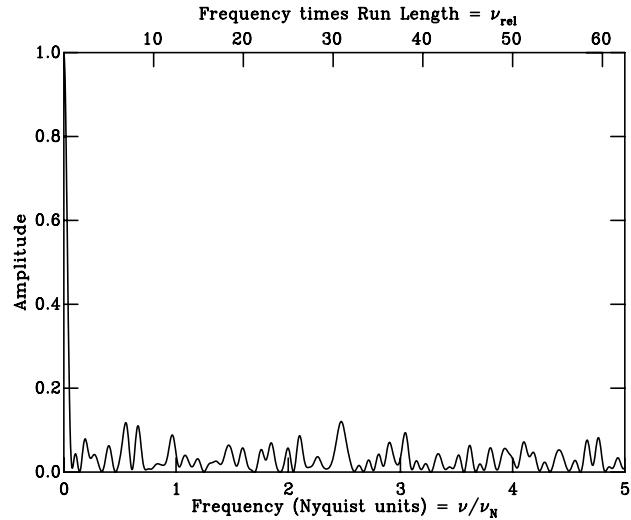}
      \caption{A typical window function for an optimum randomised geometric
      sampling of 25 data points, where $x=1.128$. The RMS is 0.025.
      \label{fig:rand_geometric_n25_wfn} }
   \end{figure}

   Figure \ref{fig:rand_geometric_n25_rms} plots the variation of RMS with
   geometric base for a 25 point reshuffled geometric spacing. For comparison,
   the ordered geometric spacing is overplotted (dashes). In general, a
   randomised geometric spacing has a slightly higher RMS than the equivalent
   ordered spacing over the range of optimal bases. Figure
   \ref{fig:rand_geometric_n25_wfn} shows the window function for the base with
   the lowest RMS. It should be noted that each random realisation has its own
   unique window function, and that for small values of $N$ the RMS values can
   vary markedly from run to run. However, the range of variation in RMS across
   the space of possible random realisations shrinks rapidly with increasing
   $N$.

   %% FIG: Reshuffled n=25 RMS plot.

   %% FIG: W. fn. for optimum reshuffled spacing, N=25.

   %% Compare ordered and randomised, using N=100 as the example.

\begin{figure}  
   \vspace{70mm}
      \includegraphics{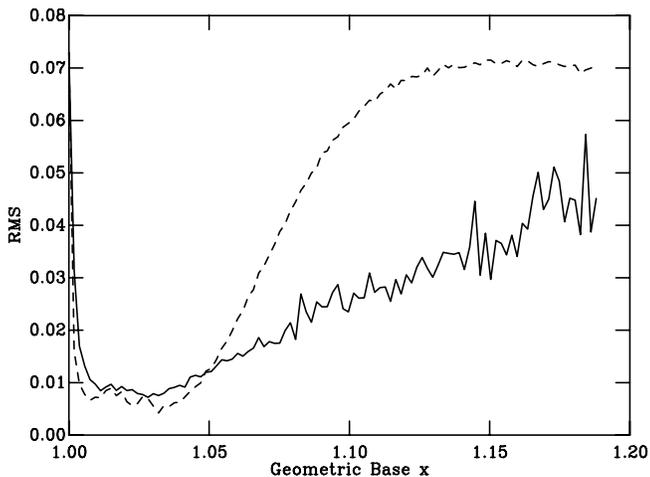}
      \caption{Randomised geometric base as a function of RMS, for $N=100$
      (solid line). For
      comparison, the ordered geometric spacing has been overplotted (dashes).
      \label{fig:rand_geometric_n100_rms} }
   \end{figure}

   The different behaviour between the ordered and randomised geometric spacing
   is more clearly illustrated in Figure \ref{fig:rand_geometric_n100_rms}. The
   optimal range of bases remains almost the same as for the ordered case. In
   general, the smoothness (as measured by the RMS value) of the window function
   of the best randomised geometric spacing is not quite as good as the
   equivalent ordered spacing. However, the randomised spacing degrades much
   more gracefully with increasing geometric base - for sub-optimal choices of
   the geometric base, a randomised spacing outperforms the ordered spacing.
   This has important implications for observing programmes in which the total
   number of observations is a highly dynamic quantity which cannot be
   accurately predicted in advance. In such a context, choosing a randomised
   geometric spacing would allow the observing programme more flexibility with
   the total number of observations and their spacings, while seeking to
   optimise the schedule based on the current estimate of the run size and
   duration. Such an algorithm could form the basis of an \emph{adaptive dataset
   planning} agent, an autonomous software entity which encapsulates sampling
   knowledge and attempts to exploit it in a real world environment.

   %% FIG: Reshuffled n=100 RMS plot.
   
\section{Practical application} \label{Section:Practical application}

   %% FIG: Optimal bases for ordered 10-500 observations plot.

   \begin{figure}  
   \vspace{70mm}
      \includegraphics{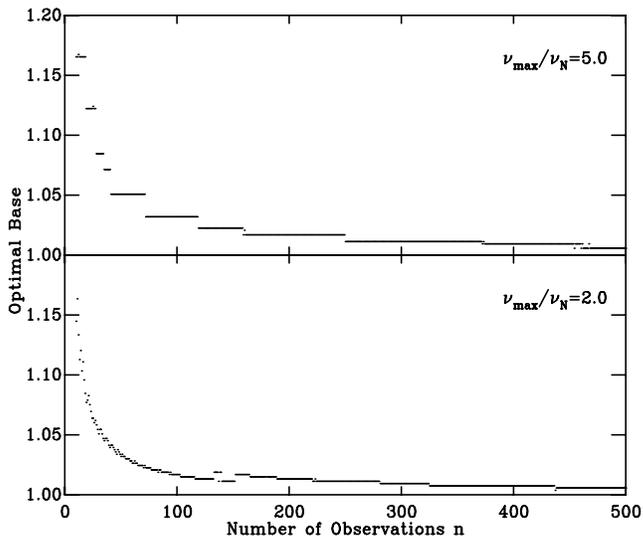}
      \vspace{5mm}
      \caption{Optimal geometric bases, as determined by minimum
      RMS, for datasets spanning 10 to 500 datapoints. For a relative maximum
      limiting frequency of 2 (lower panel), optimal bases tend to be closer to
      linear sampling than an equivalent sampling extending over a much higher
      dynamic range (upper panel). Simulations out to a relative maximum
      limiting frequency of 12.5 show almost no deviation from the optimal bases
      calculated for the $\nu_{\rm max}/\nu_{\rm N}=5$ case.      
      \label{fig:optimal_bases} }
   \end{figure}

   %% Example scenario illustrating usage of the table.

   Figure \ref{fig:optimal_bases} presents the optimal geometric base for a
   range of randomly respaced geometric sampling scenarios. By taking the period
   range and run-length of interest, converting them into a relative limiting
   frequency, and then estimating the total number of observations to be made,
   the ideal geometric base for different observation regimes can be found.
   Expressing the limiting frequency in units of $\nu_{\rm N}$, and substituting
   Equation \ref{Equation:Nyquist} gives

   \begin{equation}
   \left(\frac{\nu}{\nu_{\rm N}}\right) = \frac{2 T \nu}{N} = \frac{2\nu_{rel}}
   {N}
   \end{equation}

   As an example, an astronomer with access to a robotic network is planning a 3
   week observing run, searching for periods ranging from 5 hours to 10 days. 
   This corresponds to a minimum relative period of 0.0099, and thus a maximum
   limiting relative frequency $\nu_{rel}$ of 101. If the total number of
   observations is likely to be around 100, then $\left(\nu/\nu_{\rm N}\right)
   \approx 2$. Applying the lower curve in Figure \ref{fig:optimal_bases}, the ideal
   sampling base is found to lie at around 1.02. 
   
   Although Figure \ref{fig:optimal_bases} has, in principle, to be recalculated
   for each particular value of $\left(\nu/\nu_{\rm N}\right)$, in practice we
   find no change in the set of optimum bases for relative maximum limiting
   frequencies greater than 5, and the optimal minima are sufficiently broad
   that crude interpolation should suffice for smaller values. If the relative
   limiting frequency is found to be below 1, then enough datapoints exist
   to sample evenly - no further optimisation is required.

   Thus the observer should apply the following procedure to build an observing
   schedule. The signal range of interest must be identified, and the likely
   size of the dataset and its duration estimated. This allows the optimal
   geometric base to be calculated. Having ascertained the optimal base,
   Equation \ref{Equation:Geometric series} may be used to calculate a series of
   observation times. The resulting gaps may then be reordered as desired as the
   run proceeds, for example to accommodate practical constraints such as weather
   conditions or telescope maintenance, or to facilitate a single night's
   intensive coverage during a period of dark time.

   \section{Conclusions} \label{Section:Conclusions}

   We have found the best solution to date for placing limited numbers of
   observations in such a way as to minimise aliasing effects arising from
   undersampling. By applying a simple geometric sampling rule, we are
   able to significantly outperform previous sampling schemes. Our scheme has
   the advantage that it is easy to apply at the telescope for datasets of
   arbitrary size, has calculated empirical solutions for different observing
   scenarios, and exhibits significant flexibility arising from the invariance
   of the optimal sampling choice under the effects of reordering. As Figure
   \ref{fig:3geometric} shows, by careful sampling the improvements that can be
   made in data quality, or alternatively, the savings that may be accrued in
   telescope time are substantial.

   \begin{acknowledgements}   
   The authors would like to thank Charles Williams for many useful discussions.
   ESS is funded through a PPARC e-Science studentship. This work is part of the
   eSTAR project, which supports AA and which is jointly funded by the DTI,
   PPARC and EPSRC. We are grateful to the Starlink project for their
   flexibility in seconding AA to eSTAR. We are also grateful to the European
   Union VOTech project.
   \end{acknowledgements}

\bibliographystyle{aa}
\bibliography{ref/refs}
\end{document}